\documentclass[aps,prl,floatfix,twocolumn,footinbib]{revtex4}
\input{epsf.tex}
\usepackage[dvips]{graphicx,epsfig}
\usepackage{subfigure}
\usepackage{textcomp}
\usepackage{amsfonts}
\usepackage{amsmath}
\usepackage{amssymb}

\def\bkR{{\rm I\kern-.17em R}}
\def\bkC{{\rm \kern.24em \vrule width.05em height1.4ex depth-.05ex \kern-.26em C}}
\def\to{\rightarrow}

\def\be{\beta}

\def\frac#1#2{{\textstyle{{#1}\over {#2}}}}
\def\lsim{\mathrel{\rlap{\lower4pt\hbox{\hskip1pt$\sim$}}
    \raise1pt\hbox{$<$}}}
\def\gsim{\mathrel{\rlap{\lower4pt\hbox{\hskip1pt$\sim$}}
    \raise1pt\hbox{$>$}}}
\def\sqr#1#2{{\vcenter{\vbox{\hrule height.#2pt
         \hbox{\vrule width.#2pt height#1pt \kern#1pt
         \vrule width.#2pt}
         \hrule height.#2pt}}}}

\def\laq{\raise 0.4 ex \hbox{$<$}\kern -0.8 em\lower 0.62 ex\hbox{$\sim$}}
\def\gaq{\raise 0.4 ex \hbox{$>$}\kern -0.7 em\lower 0.62 ex\hbox{$\sim$}}

\def\be{\begin{equation}}
\def\ee{\end{equation}}
\def\ba{\begin{eqnarray}}
\def\ea{\end{eqnarray}}

\def\dalemb#1#2{{\vbox{\hrule height.#2pt
        \hbox{\vrule width.#2pt height#1pt \kern#1pt \vrule width.#2pt}
        \hrule height.#2pt}}}

\def\dalemb#1#2{{\vbox{\hrule height.#2pt
        \hbox{\vrule width.#2pt height#1pt \kern#1pt \vrule width.#2pt}
        \hrule height.#2pt}}}

\def\gtorder{\mathrel{\raise.3ex\hbox{$>$}\mkern-14mu
             \lower0.6ex\hbox{$\sim$}}}
\def\ltorder{\mathrel{\raise.3ex\hbox{$<$}\mkern-14mu
             \lower0.6ex\hbox{$\sim$}}}

\begin{document}

\leftline{November 2012}

\title{Violation of the Robertson-Schr\"odinger uncertainty principle and non-commutative quantum mechanics}

\author{Catarina Bastos\footnote{E-mail: catarina.bastos@ist.utl.pt}}

\vskip 0.3cm

\affiliation{Instituto de Plasmas e Fus\~ao Nuclear, Instituto Superior T\'ecnico \\
Avenida Rovisco Pais 1, 1049-001 Lisboa, Portugal}

\author{Orfeu Bertolami\footnote{Also at Instituto de Plasmas e Fus\~ao Nuclear, Instituto Superior T\'ecnico,
Avenida Rovisco Pais 1, 1049-001 Lisboa, Portugal. E-mail: orfeu.bertolami@fc.up.pt}}

\vskip 0.3cm

\affiliation{Departamento de F\'\i sica e Astronomia, \\
Faculdade de Ci\^encias da Universidade do Porto \\
Rua do Campo Alegre, 687,4169-007 Porto, Portugal}

\author{Nuno Costa Dias\footnote{Also at Grupo de F\'{\i}sica Matem\'atica, UL,
Avenida Prof. Gama Pinto 2, 1649-003, Lisboa, Portugal. E-mail: ncdias@meo.pt}, Jo\~ao Nuno Prata\footnote{Also at Grupo de F\'{\i}sica Matem\'atica, UL,
Avenida Prof. Gama Pinto 2, 1649-003, Lisboa, Portugal. E-mail: joao.prata@mail.telepac.pt}}

\vskip 0.3cm

\affiliation{Departamento de Matem\'{a}tica, Universidade Lus\'ofona de
Humanidades e Tecnologias \\
Avenida Campo Grande, 376, 1749-024 Lisboa, Portugal}

%\date{}

\vskip 2cm

\begin{abstract}
We show that a possible violation of the Robertson-Schr\"odinger
uncertainty principle may signal the existence of a deformation of
the Heisenberg-Weyl algebra. More precisely, we prove that any Gaussian
in phase-space (even if it violates the Robertson-Schr\"odinger
uncertainty principle) will always be a quantum state of an
appropriate non-commutative extension of quantum mechanics.
Conversely, all canonical non-commutative extensions of quantum
mechanics display states that violate the Robertson-Schr\"odinger
uncertainty principle.
\end{abstract}

\maketitle

\section{Introduction}

Since its inception, quantum mechanics was formulated in terms of
Hilbert spaces and self-adjoint operators acting therein. In this
context Heisenberg's uncertainty relations become a
straightforward consequence of the non-commutativity of the
fundamental operators of position $\widehat{Q}$ and momentum
$\widehat{P}$ and the Cauchy-Schwarz inequality. For a
$n$-dimensional system, the Heisenberg-Weyl (HW) algebra reads:
\begin{equation}
\left[\widehat{Q}_i, \widehat{P}_j \right] = i \delta_{i,j} , \hspace{1 cm} i,j=1, \cdots, n,
\label{eqHeisenbergAlgebra1}
\end{equation}
and all remaining commutators vanish. Since our results are much
more general than simple rescalings of Planck's constant, we have
set $\hbar=1$ for the remainder of this work and assumed that
position and momentum have the same units. If we define the
phase-space variable $\widehat{Z} =(\widehat{Q}, \widehat{P})$, we
have in more compact notation
\begin{equation}
\left[\widehat{Z}_i, \widehat{Z}_j \right] = i J_{ij} , \hspace{1 cm} i,j=1, \cdots, 2n,
\label{eqHeisenbergAlgebra2}
\end{equation}
where ${\bf J} = \left(J_{ij} \right)$ is the standard symplectic matrix
\begin{equation}
{\bf J} = - {\bf J}^T = - {\bf J}^{-1} = \left(
\begin{array}{c c}
{\bf 0} & {\bf I}\\
- {\bf I} & {\bf 0}
\end{array}
\right).
\label{eqStandardSympMatrix}
\end{equation}
A simple calculation leads to the inequalities
\begin{equation}
\Delta_{Q_j} \cdot \Delta_{P_j} \ge \frac{1}{2}, \hspace{1 cm} j=1, \cdots, n,
\label{eqHeisenbergUncertaintyPrinciple}
\end{equation}
where $\Delta_{Q_j}$ and $\Delta_{P_j}$ denote the mean standard-deviations of $\widehat{Q}_j$ and $\widehat{P}_j$ with respect to an arbitrary state. The set of inequalities (\ref{eqHeisenbergUncertaintyPrinciple}) are known as the Heisenberg-Weyl-Pauli inequalities. They are not invariant under linear symplectic transformations of the operators nor under metaplectic transformations of the states. This prompted the search for an alternative set of inequalities which are stronger than the Heisenberg inequalities and have the right symplectic covariance properties. They are known as the Robertson-Schr\"odinger uncertainty principle (RSUP) \cite{Robertson}:
\begin{equation}
{\bf \Sigma} + \frac{i}{2} {\bf J} \ge 0.
\label{eqRobertsonSchrodingerUncertaintyPrinciple}
\end{equation}
Here ${\bf \Sigma}$ denotes the covariance matrix with respect to
an arbitrary state and has entries $\Sigma_{ij} = < \left(
\widehat{Z}_i - <\widehat{Z}_i> \right) \left( \widehat{Z}_j -
<\widehat{Z}_j> \right)>$ $(i,j =1 , \cdots, 2n)$ and $< \cdot>$
denotes an expectation value in a chosen state. In this work we
shall focus on this form of the uncertainty principle, as it
implies the Heisenberg uncertainty relations, it is invariant
under linear symplectic transformations and moreover it
constitutes the necessary and sufficient condition for a Gaussian
to be a quantum mechanical state. In fact, Gaussians are of the
utmost importance, not only because they are completely determined
by their covariance matrix, but also because experimentally
coherent and squeezed states play an important role in quantum
optics \cite{Simon}, quantum computation of continuous variables
\cite{Giedke} and investigations of the quantum-classical
transition \cite{Littlejohn}.

From a slightly different perspective, since the RSUP accounts for
correlations, it is more suitable to address several interesting
problems. The recent work \cite{Adamenko} presents one such
example. It is shown that the consideration of states with strong
position-momentum correlations may lead to greater transparency of
the Coulomb barrier during the interaction of charged particles.
This is very important in the astrophysics of stars and in
controlled nuclear fusion, where the action of the Coulomb barrier
leads to a very low tunneling probability for low-energy
particles. It can be shown that for a nonstationary harmonic
oscillator with potential $V(t) = \frac{1}{2} m \omega^2 (t) Q^2$,
a decrease in particle's frequency $\omega(t)$ leads to an
increase in the correlation coefficient:
$$
r(t) = \frac{< \widehat{Q}\widehat{P} + \widehat{P}\widehat{Q}>}{2 \Delta_Q \Delta_P}
$$
and a change of the uncertainty relation
$$
\Delta_Q \Delta_P \ge (2 \sqrt{1 - r^2})^{-1}.
$$
When a strongly correlated state with $|r| \to 1$ is formed, both
$\Delta_Q, \Delta_P$ and their product increase indefinitely. It
was then proved that this leads to a greater barrier transparency
at the same energy \cite{Adamenko}.

Various authors have tested theoretically and experimentally the
validity of the inequalities
(\ref{eqRobertsonSchrodingerUncertaintyPrinciple}) and the
consequences of their violation \cite{Rozema}. For instance,
Popper's experiment is usually regarded as a violation of the
uncertainty principle \cite{Kim}. The type of arguments used by
Popper \cite{Popper} are similar in spirit to those of the EPR
experiment \cite{Einstein} so that non-locality and the
uncertainty principle seem to be inextricably linked. More
precisely, the degree of non-locality of any theory is determined
by two factors, namely the strength of the uncertainty principle, and the
strength of a property called ``steering", which determines which
states can be prepared at one location given a measurement at
another \cite{Oppenheim}.

In this work, we show that the breakdown of the uncertainty
principle may hint that the sub-atomic world is described not by
standard quantum mechanics but by a non-commutative extension
\cite{Bastos1,Bertolami} of it. This extension is obtained by
replacing HW algebra (\ref{eqHeisenbergAlgebra2}) by a {\it
deformed} algebra. This leads to an extra non-commutativity
between the configuration and momentum variables. The most
commonly used deformed algebra reads:
\begin{equation}
\left[\widehat{\Xi}_i, \widehat{\Xi}_j \right] = i \Omega_{ij}, \hspace{1 cm} i,j = 1, \cdots, 2n,
\label{eqDeformedHeisenbergAlgebra}
\end{equation}
where the matrix ${\bf \Omega} = \left( \Omega_{ij} \right)$ is given by:
\begin{equation}
{\bf \Omega} = \left(
\begin{array}{c c}
{\bf \Theta} & {\bf I}\\
- {\bf I} & {\bf \Upsilon}
\end{array}
\right)
\label{eqDeformedSymMatrix}
\end{equation}
Here ${\bf \Theta}=(\theta_{ij})$ and ${\bf \Upsilon}=(\eta_{ij})$ are real constant skew-symmetric $n \times n$ matrices measuring the strengths of the position-position and momentum-momentum non-commutativities, respectively.

One should notice that other (non-canonical) deformations of the
algebra (\ref{eqHeisenbergAlgebra2}) are also possible. These
typically account for non-linearities and may appear in various
systems. For instance the quantum $q$-oscillator \cite{Biedenharn}
can be interpreted as a nonlinear oscillator whose
frequency of vibration depends on the energy of the vibrations
through a $\cosh$ function \cite{Manko1}. Generalizations thereof, where the
frequency of vibration varies with other functions of the
amplitude, are commonly known as $f$-oscillators \cite{Manko2}.
Another instance where this kind of algebraic structure is common
is in the reduced phase space formulation of systems with second
class Dirac constraints \cite{Henneaux}.

From a somehow more fundamental point of view, non-commutative
quantum mechanics (NCQM) is usually regarded as a non-relativistic
one-particle sector of the very discussed non-commutative quantum
field theories \cite{Douglas}, which emerge in the context of
string theory and quantum gravity \cite{Connes}. The question of
space-time non-commutativity has a long-standing story. It was put
forward by Snyder \cite{Snyder}, Heisenberg, Pauli
\cite{Heisenberg} and Yang \cite{Yang} as a means to regularize
quantum field theories. However, the development of
renormalization techniques and certain undesirable features of
noncommutative theories such as the breakdown of Lorentz 
invariance \cite{Carrol,Bertolami-Guisado} have hindered further research in this direction. More
recently, several important developments in various approaches to
the quantization of gravity have revived the interest in the
concept of noncommutative space-time. See for instance
Ref. \cite{Freidel} in the context of 3d gravity. In the realm of
string theory, the discovery that the low energy effective theory
of a D-brane in the background of a Neveu-Schwarz B field lives on
a space with spatial non-commutativity  has triggered an enormous
amount of research in this field \cite{Douglas,Douglas2,Connes}.
From another perspective, a simple heuristic argument, based on
Heisenberg's uncertainty principle, the equivalence principle and
the Schwarzschild metric, shows that the Planck length seems to be
a lower bound on the precision of a measurement of position
\cite{Rosenbaum}. This reenforces the point of view that a new
(non-commutative) geometry of space-time will emerge at a
fundamental level \cite{Douglas,Connes2,Martinetti,Szabo}. The
simplest way to implement these ideas in quantum field theories is
by adding to the phase-space non-commutativity of quantum
mechanics a new space-time non-commutativity.

The implications of considering NC deformations of the HW algebra
have also been investigated in the context of quantum cosmology.
It was shown that this new structure leads to the thermodynamic
stability of black holes and a possible regularization of the
black hole singularities \cite{Bastos3}.

In this paper, we study some fundamental properties of states in
NCQM. Various features of noncommutative quantum states for the
algebra
(\ref{eqDeformedHeisenbergAlgebra})-(\ref{eqDeformedSymMatrix})
have already been derived \cite{Bastos1}. Here we are concerned
with the possibility of using Gaussian states to identify the
correct algebraic structure for quantum mechanics. More precisely,
we show that: (i) NCQM displays states which are not states in
standard quantum mechanics (as they violate the RSUP); (ii)
conversely, that any Gaussian state (even if it is not a state in
standard quantum mechanics) is  nevertheless a state of some NC
quantum theory; (iii) this NC theory is not unique and one can
always find a NCQM for which the Gaussian is a quantum pure state.
We stress that that all these results are true independently of
the value of Planck's constant and we set $\hbar=1$ for the entire
paper.

Our proof takes place in the context of the Weyl-Wigner
formulation or deformation quantization framework, where position
and momentum variables appear on equal footing. As shown in
\cite{Bastos1}, it is only in this formulation that one can tell
whether states correspond to a quantization of the HW algebra or
some of its deformation.

\section{Wigner distributions, and Gaussian states on arbitrary symplectic spaces}

A symplectic form on a real vector space $V$ is a bilinear map
$\omega: V \times V \to \mathbb{R}$, which is skew-symmetric
($\omega (v, v^{\prime}) = - \omega (v^{\prime}, v)$ for all
$v,v^{\prime} \in V$) and non-degenerate ($\omega (v, v^{\prime})
=0$ for all $v^{\prime} \in V$ implies $v=0$). The archetypal
symplectic vector space is $V= \mathbb{R}^{2n}= \mathbb{R}^n_x
\times \mathbb{R}^n_p$ endowed with the standard symplectic form:
\begin{equation}
\sigma (z,z^{\prime}) = z \cdot {\bf J} z^{\prime} = p \cdot x^{\prime} - x \cdot p^{\prime},
\label{eqStandSymForm}
\end{equation}
where ${\bf J}$ is the standard symplectic matrix and $z = (x,p), ~ z^{\prime} = (x^{\prime},p^{\prime}) \in \mathbb{R}^{2n}$.

Now, let
\begin{equation}
\omega (z, z^{\prime}) = z \cdot {\bf \Omega}^{-1} z^{\prime}
\label{eqNonStandSympForm}
\end{equation}
be another arbitrary symplectic form on $\mathbb{R}^{2n}$. Here
${\bf \Omega}$ is some real, anti-symmetric, non-singular, $2n
\times 2n$ matrix (not necessarily (\ref{eqDeformedSymMatrix})). A
well known theorem in symplectic geometry \cite{Cannas} states
that all symplectic vector spaces of equal dimension are
symplectically equivalent. In other words, there exists a real
non-singular $2n \times 2n$ matrix ${\bf S}$ such that $\omega
({\bf S} z, {\bf S} z^{\prime}) = \sigma( z, z^{\prime})$ for all
$z,z^{\prime} \in \mathbb{R}^{2n}$, or matrix-wise
\begin{equation}
{\bf \Omega} = {\bf S} {\bf J} {\bf S}^T.
\label{eqDarbouxMatrix}
\end{equation}
Obviously, the matrix ${\bf S}$ is not unique. Indeed, let ${\bf P}$ denote a symplectic matrix $({\bf P} \in Sp (2n; \sigma))$, that is ${\bf P} {\bf J} {\bf P}^T = {\bf J}$. Then the matrix ${\bf S}{\bf P}$ also satisfies (\ref{eqDarbouxMatrix}). We shall call the set of all matrices which satisfy (\ref{eqDarbouxMatrix}) the set of Darboux matrices associated with $ \omega$ and denote it by $\mathcal{D}(2n; \omega)$.

Conversely, given any real, non-singular, $2n \times 2n$ matrix ${\bf S}$, let ${\bf \Omega}$ be defined by (\ref{eqDarbouxMatrix}). Then $\omega$ given by (\ref{eqNonStandSympForm}) is a symplectic form on $\mathbb{R}^{2n}$. This simple observation will be the crux of our main result.

\subsection{Quantization on the standard symplectic space}

To quantize a system on the standard symplectic space, one resorts
to the HW operators
\begin{equation}
\widehat{U}_{\sigma}(z) = e^{i \sigma(z , \widehat{Z})} = e^{i (p \cdot \widehat{Q} - x \cdot \widehat{P})}
\label{eqWeylOperators}
\end{equation}
with $z=(x,p) \in \mathbb{R}^{2n}$, and where $\widehat{Z} = (\widehat{Q}, \widehat{P})$ denote the quantum mechanical position and momentum operators.

These operators constitute a unitary irreducible representation of the HW algebra. Indeed, they satisfy the relations
\ba
\widehat{U}_{\sigma}(z) \widehat{U}_{\sigma}(z^{\prime}) &=& e^{\frac{i}{2} \sigma (z, z^{\prime})} \widehat{U}_{\sigma}(z + z^{\prime}) \nonumber\\
&=& e^{i \sigma (z, z^{\prime})} \widehat{U}_{\sigma}(z^{\prime}) \widehat{U}_{\sigma}(z),
\label{eqWeylAlgebra}
\ea
which can be readily obtained from the HW algebra (\ref{eqHeisenbergAlgebra2}) through the Baker-Campbell-Hausdorff formula.

A generic linear operator acting on the Hilbert space of the
system (in the case $L^2 (\mathbb{R}^n)$) can then be represented
by
\begin{equation}
\widehat{A} =  (2 \pi )^{-n} \int dz ~ \widetilde{\alpha}_{\sigma} ({\bf J}^{-1} z) \widehat{U}_{\sigma} (z) ,
\label{eqOperator}
\end{equation}
where $\widetilde{\alpha}_{\sigma}$ denotes the Fourier transform of some suitable tempered distribution $\alpha_{\sigma} (z)$ on the phase-space, commonly known as the Weyl-symbol of the operator $\widehat{A}$.

Since $Tr (\widehat{U}_{\sigma} (z))= (2 \pi)^n \delta (z)$, we obtain from (\ref{eqWeylAlgebra}) that $Tr (\widehat{U}_{\sigma} (z) \widehat{U}_{\sigma} (z^{\prime})) = (2 \pi)^n \delta (z + z^{\prime})$, and (\ref{eqOperator}) can be readily inverted. Thus, $\widetilde{\alpha}_{\sigma} (z) =Tr \left( \widehat{A} \widehat{U}_{\sigma} ({\bf J}^{-1} z) \right)$, and the Weyl symbol of $\widehat{A}$ reads:
\ba
&&\widehat{A} \mapsto W_{\sigma} \widehat{A} = \alpha_{\sigma} (z) \nonumber\\
&&=  (2 \pi )^{-n} \int d z^{\prime} ~  Tr \left( \widehat{A} \widehat{U}_{\sigma} (z^{\prime}) \right) e^{i \sigma (z^{\prime},z)}  \nonumber\\
&&= \int dy ~< x + \frac{y}{2}| \widehat{A} | x - \frac{y}{2}> e^{- i p \cdot y} .
\label{eqWeylSymbol}
\ea
This procedure establishes a one-to-one map $W_{\sigma}$ - called Weyl correspondence - between linear operators and the associated symbols.

The state of a quantum system is represented by a positive trace-class operator, the density matrix $\widehat{\rho}$. When applied to $\widehat{\rho}$ the Weyl correspondence yields (up to a normalization constant) the celebrated Wigner function on $(\mathbb{R}^{2n} , \sigma)$:
\begin{equation}
W_{\sigma} \widehat{\rho} (x,p) =  (2 \pi )^{-n} \int dy ~< x + \frac{y}{2}| \widehat{\rho} | x - \frac{y}{2}> e^{- i p \cdot y} .
\label{eqWignerFunction}
\end{equation}
In particular, for a pure state $\widehat{\rho} = | \psi>< \psi|$ for $\psi \in L^2 (\mathbb{R}^n)$:
\begin{equation}
W_{\sigma} \psi (x,p) = (2 \pi)^{-n} \int dy ~ \psi \left(x + \frac{y}{2}\right) \overline{ \psi \left(x - \frac{y}{2}\right)} e^{- i p \cdot y}.
\label{eqWignerFunctionPureState}
\end{equation}
In general, it is very difficult to assess whether a function $F(x,p)$ in phase-space is the Wigner function of some density matrix (see e.g. \cite{Dias2}). A notable exception are the Gaussians
\begin{equation}
\mathcal{G}_{{\bf \Sigma} , \zeta} (z) = \frac{1}{(2 \pi)^n\sqrt{\det {\bf \Sigma}}} \exp \left[ - \frac{1}{2} (z- \zeta) \cdot {\bf \Sigma}^{-1} (z- \zeta)\right]
\label{eqGaussian}
\end{equation}
where $z=(x,p) \in \mathbb{R}^{2n}$, $\zeta \in \mathbb{R}^{2n}$ and ${\bf \Sigma}$ is the covariance matrix, which is a real, positive-definite $2n \times 2n $ matrix. It is a well documented fact \cite{Simon,Dias2} that such a Gaussian is a Wigner function on $(\mathbb{R}^{2n} ; \sigma)$ if and only if it satisfies the RSUP (\ref{eqRobertsonSchrodingerUncertaintyPrinciple}).

We may be more specific and determine whether the Gaussian is the Wigner function of a pure state. Indeed, it has been proven \cite{Littlejohn} that (\ref{eqGaussian}) is the Wigner function of a pure state if and only if there exists ${\bf P} \in Sp (2n; \sigma)$ such that
\begin{equation}
{\bf \Sigma} = \frac{1}{2} {\bf P}^T {\bf P}.
\label{eqLittlejohnTheorem}
\end{equation}

\subsection{Quantization on non-standard symplectic spaces}

NCQM results from quantizing the classical theory on a
non-standard symplectic space $(\mathbb{R}^{2n}; \omega)$
\cite{Bastos1,Bertolami}. In this case, the HW algebra
(\ref{eqHeisenbergAlgebra2}) is replaced by the modified algebra
(\ref{eqDeformedHeisenbergAlgebra}). The HW operators
(\ref{eqWeylOperators}) become \be \widehat{U}_{\omega} (\xi )=
e^{i \omega \left(\xi, \widehat{\Xi} \right)} = e^{ i  \xi \cdot
{\bf \Omega}^{-1} \widehat{\Xi}}. \label{eqModifiedWeylOperators}
\ee and concomitantly \ba
\widehat{U}_{\omega} (\xi )\widehat{U}_{\omega} (\xi^{\prime} ) &=& e^{\frac{i}{2} \omega(\xi, \xi^{\prime})} \widehat{U}_{\omega} (\xi + \xi^{\prime} )\nonumber\\
& =& e^{i \omega (\xi, \xi^{\prime})} \widehat{U}_{\omega} (\xi^{\prime} ) \widehat{U}_{\omega} (\xi).
\label{eqModifiedWeylAlgebra}
\ea
Since $\widehat{\Xi} = {\bf S} \widehat{Z}$ for some ${\bf S} \in \mathcal{D} (2n; \omega)$, from (\ref{eqDarbouxMatrix}):
\ba
\widehat{U}_{\omega} (\xi) &=& \widehat{U}_{\sigma} ({\bf S}^{-1} \xi)~,\nonumber\\
Tr (\widehat{U}_{\omega} (\xi)\widehat{U}_{\omega} (\xi^{\prime})) &=&  (2 \pi)^n \sqrt{\det {\bf \Omega}}\delta  (\xi + \xi^{\prime}).
\label{eqWeylModifiedOperator}
\ea
Substituting (\ref{eqWeylModifiedOperator}) into (\ref{eqOperator}), we obtain
\begin{equation}
\widehat{A} = (2 \pi \sqrt{\det {\bf \Omega}})^{-n} \int dz^{\prime} ~\widetilde{a}_{\omega} ({\bf \Omega}^{-1} z^{\prime}) \widehat{U}_{\omega} (z^{\prime}),
\label{eqModifiedWeylOperator}
\end{equation}
where (\ref{eqDarbouxMatrix}) and $\widetilde{a}_{\omega} (u) = \widetilde{a}_{\sigma} ({\bf S}^T u)$ have been used. Hence, the Weyl symbol of $\widehat{A}$ on $(\mathbb{R}^{2n}; \omega)$ is given by
\begin{equation}
\begin{array}{c}
\widehat{A} \mapsto W_{\omega} \widehat{A} = a_{\omega} (\xi) = (\sqrt{\det{\bf \Omega}})^{-1}  a_{\sigma} ({\bf S}^{-1} \xi) =\\
\\
=((2 \pi)^n \det {\bf \Omega})^{-1} \int d \xi^{\prime} ~ Tr \left(\widehat{A} \widehat{U}_{\omega} (\xi^{\prime}) \right) e^{i \omega (\xi^{\prime}, \xi)}.
\end{array}
\label{eqModifiedWeylSymbol}
\end{equation}
In particular if $W_{\sigma} \widehat{\rho} $ denotes the Wigner function of a density matrix $\widehat{\rho}$ on $(\mathbb{R}^{2n}; \sigma)$, then the corresponding Wigner function on $(\mathbb{R}^{2n}; \omega)$ is given by \cite{Bastos1}
\begin{equation}
W_{\omega} \widehat{\rho} (\xi) = \frac{1}{\sqrt{ \det {\bf \Omega}}} W_{\sigma} \widehat{\rho} ({\bf S}^{-1} \xi).
\label{eqModifiedWignerFunction}
\end{equation}

\subsection{Main result}

From (\ref{eqModifiedWignerFunction}) one can derive the
counterparts of the RSUP and Littlejohn's Theorem
\cite{Littlejohn} for Gaussians on $(\mathbb{R}^{2n}; \omega)$.
From (\ref{eqRobertsonSchrodingerUncertaintyPrinciple}) and
(\ref{eqModifiedWignerFunction}), we conclude that the Gaussian
(\ref{eqGaussian}) is a quantum state on $(\mathbb{R}^{2n};
\omega)$ if and only if
\begin{equation}
{\bf \Sigma} + \frac{i}{2} {\bf \Omega} \ge 0.
\label{eqModifiedRobertsonSchrodingerUncerPrinc}
\end{equation}
Likewise, from (\ref{eqLittlejohnTheorem}) and (\ref{eqModifiedWignerFunction}), the Gaussian is a Wigner function on $(\mathbb{R}^{2n}; \omega)$ of a pure state if and only if there exists a matrix ${\bf C} \in \mathcal{D} (2n; \omega)$ such that
\begin{equation}
{\bf \Sigma} = \frac{1}{2} {\bf C} {\bf C}^T.
\label{eqModifiedLittlejohnTheorem}
\end{equation}

 We are now in condition prove our main result.

\begin{center}
{\it Let ${\mathcal{G}_{{\bf \Sigma} ,  \zeta}}$ be a Gaussian of
the form (\ref{eqGaussian}). Then there exists a matrix ${\bf
\Omega}$ associated with a symplectic form
(\ref{eqNonStandSympForm}), such that ${\mathcal{G}_{{\bf \Sigma}
, \zeta}}$ is a quantum state of the NCQM based on the deformed
Heisenberg algebra (\ref{eqDeformedHeisenbergAlgebra}).}
\end{center}

Indeed, since ${\bf \Sigma}$ is positive definite, there exists a real, non-singular, $2n \times 2n$ matrix ${\bf C}$ for which (\ref{eqModifiedLittlejohnTheorem}) holds. Define a matrix ${\bf \Omega}$ by ${\bf \Omega} = {\bf C} {\bf J} {\bf C}^T$. Clearly, the form $\omega$ defined by (\ref{eqNonStandSympForm}) is a symplectic form and ${\bf C} \in \mathcal{D} (2n; \omega)$. According to Littlejohn's Theorem on $(\mathbb{R}^{2n}; \omega)$ (cf.(\ref{eqModifiedLittlejohnTheorem})), then ${\mathcal{G}_{{\bf \Sigma} ,  \zeta}}$ is a Wigner function on $(\mathbb{R}^{2n}; \omega)$ of a pure state. This concludes the proof.

Notice that, as a by-product of the proof, for any Gaussian, one
can always find a NCQM for which the Gaussian is a pure state. In
a certain sense this procedure amounts to a purification of the
state.

There is a converse result of the previous one. Namely: given a
NCQM we can always find quantum states which violate the standard
RS uncertainty principle. A general proof of this result is beyond
the scope of the present letter. Here we shall illustrate
explicitly this result for NCQM in 2 dimensions with algebra
(\ref{eqDeformedHeisenbergAlgebra})-(\ref{eqDeformedSymMatrix}).
That is
\begin{equation}
{\bf \Omega} = \left(
\begin{array}{c c}
\theta {\bf E} & {\bf I}\\
- {\bf I} & \eta {\bf E}
\end{array}
\right), \hspace{1 cm} {\bf E} = \left(
\begin{array}{c c}
0 & 1\\
-1 & 0
\end{array}
\right),
\label{eqDeformedSymMatrix1}
\end{equation}
where $\theta, \eta >0$ are real constants such that $\xi = \theta \eta <1$. A simple Darboux matrix is ${\bf C} \in \mathcal{D} (2n; \omega)$ given by
\begin{equation}
{\bf C} =
\left(
\begin{array}{c c}
\lambda {\bf I} & - \frac{\theta}{2 \lambda} {\bf E}\\
\frac{\eta}{2 \mu} {\bf E} & \mu {\bf I}
\end{array}
\right)
\label{eqDarbouxMatrix1}
\end{equation}
where $\mu, \lambda$ are real parameters such that $2 \mu \lambda
= 1 + \sqrt{1- \xi}$. By construction the Gaussian with covariance
matrix ${\bf \Sigma} = \frac{1}{2} {\bf C}{\bf C}^T$ is a Wigner
function on the non-standard symplectic space $(\mathbb{R}^4,
\omega)$. However, a straightforward calculation reveals that it
violates the standard RSUP
(\ref{eqRobertsonSchrodingerUncertaintyPrinciple}).

\section{Conclusions}

Let us close our discussion with some clarifying remarks. The
symplectic form $\omega$ such that ${\mathcal{G}_{{\bf \Sigma} ,
\zeta}}$ is a quantum state on $(\mathbb{R}^{2n}; \omega)$ is not
unique. Indeed, a phase-space function may be a Wigner function
upon quantization on several distinct symplectic spaces (see
\cite{Bastos1} for examples). Moreover, although our result proves
that a Gaussian will always be associated with some pure state on
an appropriate symplectic space, it may nevertheless be a Wigner
function on another symplectic space, this instance associated
with a mixed state.

Finally, one could wonder whether any {\it reasonable} function on phase-space will always be a Wigner function on some appropriate symplectic space $(\mathbb{R}^{2n}; \omega)$. By a {\it reasonable} function, we mean a function which satisfies some obvious {\it a priori} requirements of Wigner functions: being real, normalized, uniformly continuous, bounded, square integrable, etc. The answer is negative. Indeed if a function $F(z)$ is a Wigner function on some $(\mathbb{R}^{2n}; \omega)$, then there has to exist a matrix ${\bf S}$ such that $| \det {\bf S}| F({\bf S}z)$ is a Wigner function on $(\mathbb{R}^{2n}; \sigma)$. But one may easily construct functions which are {\it reasonable} but are not Wigner functions on $(\mathbb{R}^{2n}; \sigma)$, even admitting rescalings of Planck's constant. The derivation of such examples is difficult and beyond the scope of the present work. It requires the concept of Narcowich-Wigner spectrum \cite{Dias2}. We shall get back to this issue in a future work.

\begin{acknowledgements}

The work of CB is supported by Funda\c{c}\~{a}o para a Ci\^{e}ncia
e a Tecnologia (FCT) under the grant SFRH/BPD/62861/2009. The work
of OB is partially supported by the FCT project
PTDC/FIS/111362/2009. N.C. Dias and J.N. Prata have been supported
by the FCT grant PTDC/MAT/099880/2008.

\end{acknowledgements}

\end{document}